# Generation of stable multi-vortex clusters in a dissipative medium with anti-cubic nonlinearity

Yunli Qiu[1], Boris A. Malomed[2], Dumitru Mihalache[3], Xing Zhu[1], Jianle Peng[1], and Yingji He[1]*

[1] *School of Photoelectric Engineering, Guangdong Polytechnic Normal University, Guangzhou 510665, China*

[2] *Faculty of Engineering, Department of Physical Electronics, School of Electrical Engineering, Tel Aviv University, Tel Aviv 69978, Israel*

[3] *Horia Hulubei National Institute for Physics and Nuclear Engineering, P.O. Box MG-6, RO-077125, Bucharest-Magurele, Romania*

*Corresponding author: heyingji8@126.com

We demonstrate the generation of vortex solitons in a model of dissipative optical media with thesingular anti-cubic (AC) nonlinearity, by launching a vorticity-carrying Gaussian input into the medium modeled by the cubic-quintic complex Ginzburg-Landau equation. The effect of the AC term on the beam propagation is investigated in detail. An analytical result is produced for the asymptotic form of fundamental and vortical solitons at the point of $r \to 0$, which is imposed by the AC term. Numerical simulations identify parameter domains that maintain stable dissipative solitons in the form of vortex clusters. The number of vortices in the clusters is equal to the vorticity embedded in the Gaussian input.

# 1. INTRODUCTION

Dynamics of spatial optical solitons in conservative and dissipative mediais a vast research area,with a potential for applications to all-optical switching and light steering, pattern recognition, parallel data processing, *etc.* [1–9].Universal models for the light propagation in dissipative settings are provided by the complex Ginzburg-Landau (CGL) equations, which find well-known realizations in many areas, including superconductivity and superfluidity, fluid dynamics, reaction-diffusion pattern formation, nonlinear optics, Bose–Einstein condensates, quantum-field theories, and biology [10,11]. In particular, the CGL equations serve as realistic dynamical models of laser cavities, which admit the formation of stable fundamental and vortex solitons, as well as multi-soliton clusters [12–24].

Vortex solitons of the dark type, supported by an optical background field, represent an important class of topologically organized patterns in defocusing nonlinear media [25,26]. They carry a spiral phase structure, with zero amplitude at the center [27]. In general, only vortices with the unitary topological charge are stable, their propagation in optical waveguides being limited by a finite extension of the background field, which limits applications of dark vortex solitons [25-33]. The basic applications include the transfer of the optical angular momentum from light to matter [34], guiding light by light [35], and manipulation of nanoparticles in colloidal suspensions [36]. In turn, the dynamics of vortex solitons can be steered by inhomogeneity of the intensity and phase of the carrying background field [37–41].

Ring-shaped vortex solitons of the bright type, with zero background, have also been predicted in media with self-focusing or more complex self-defocusing nonlinearities [42,43]. In particular, the creation of robust multi-vortex clusters embedded in two-dimensional beams [44], and ring-shaped optical vortices in media described by the generic cubic-quintic (CQ) CGL equation with an inhomogeneous effective diffusion term was studied recently by us [45,46].

A new form of nonlinearity in CGL equations, in the form of the *anti-cubic* (AC) term,$\sim |u|^{-4}u$ (see Eq. (1) below), was first introduced in 2003 [47]. Since then, many works have considered the use of this nonlinearity for the generation and steering of solitons, including applications such as cooling of optical solitons [48], creation of solitons in metamaterials [49], and control of solitons in higher dimensions [50].

In this work, we report the generation of vortex solitons in a model of dissipative optical media combining the CQ and AC nonlinearities. The model is introduced in Section 2, which also includes an analytical result for a specific asymptotic form of fundamental and vortical solitons in the model including the AC term. Numerical results, produced by systematic simulations of the evolution of Gaussian-shaped inputs, which carry the angular momentum, are reported in Section 3. The results demonstrate that stable single and multiple vortices are readily produced in such media. In particular, we conclude that the number of vortices in the multiple clusters is equal to the vorticity embedded in the input Gaussian. The paper is concluded by Section 4.

## 2. THE MODEL AND ANALYTICAL RESULTS

Starting from models of the CGL type with the effective diffusion and CQ

nonlinearity [15–22, 44, 45], and adding the AC term, we adopt the following propagation equation in the spatial domain:

$$iu_z + (1/2)\Delta u + |u|^2 u - \nu|u|^4 u + Q|u|^{-4}u = iR[u], \quad (1)$$

where $\Delta = \partial^2/\partial x^2 + \partial^2/\partial y^2$ is the transverse Laplacian ($x$ and $y$ are the transverse coordinates, and $z$ is the propagation distance), the coefficient in front of the cubic self-focusing term is scaled to be 1, $\nu \geq 0$ is the quintic self-defocusing coefficient, and $Q$ is the coefficient of the AC nonlinearity [47-49].

Phenomenologically, the propagation equation including the AC term may apply to the light propagation in a medium in which the linear propagation is blocked due to a strong resonance of the electromagnetic wave with a material resonance, while a finite amplitude of the wave detunes the resonance, making the propagation possible, with a singular effective refractive index, effectively represented by the AC term in Eq. (1). It is relevant to mention that the singularity somewhat resembles terms known in models for surface waves in plasmas confined by a sharp boundary [51-53].

Loss and gain terms are collected on the right-hand side of Eq. (1),

$$R[u] = -\delta u + \varepsilon|u|^2 u - \mu|u|^4 u + \widehat{D}u, \quad (2)$$

where $\delta > 0$ is the linear-loss coefficient, $\mu > 0$ is the quintic-loss parameter, and $\varepsilon > 0$ is the cubic-gain coefficient. The last term represents the effective diffusion in the usual form:

$$\widehat{D}u = \beta\left[\frac{\partial^2 u}{\partial x^2} + \frac{\partial^2 u}{\partial y^2}\right], \quad (3)$$

with the respective positive coefficient $\beta$.

To solve Eq. (1), we first consider the equation including the linear and AC terms,

$$iu_z + i\delta u + \left(\frac{1}{2} - i\beta\right)\Delta u + Q|u|^{-4}u = 0, \quad (4)$$

and look for vortex solutions in polar coordinates $(r, \theta)$ as

$$u(z, r, \theta) = U(r) \exp(ikz + iM\theta), \quad (5)$$

where $k$ is the real propagation constant, $M$ is the vorticity (topological charge of the input optical beam), and the complex function $U(r)$ satisfies the equation

$$(k - i\delta)U = \left(\frac{1}{2} - i\beta\right)\left(\frac{d^2U}{dr^2} + \frac{1}{r}\frac{dU}{dr} - \frac{M^2}{r^2}U\right) + Q|U|^{-4}U. \quad (6)$$

In the absence of the AC term ($Q = 0$), the asymptotic form of solutions to Eq. (6) at $r \to 0$ is commonly known as $U(r) = U_0 r^M$, with arbitrary constant $U_0$. In the presence of the AC term, the asymptotic form of the solution changes to

$$U(r) = U_0 r^{\frac{1}{2} + i\lambda}, \quad (7)$$

with real amplitude $U_0$ and *chirp* $\lambda$. Substitution of ansatz (7) in Eq. (6) and keeping in it terms that are most singular at $r \to 0$, yields the following result:

$$\left(\frac{1}{2} - i\beta\right)\left[M^2 - \left(\frac{1}{2} + i\lambda\right)^2\right] = QU_0^{-4}. \quad (8)$$

The condition of the reality of the expression on the left-hand side of Eq. (8) leads to a quadratic equation for the chirp, $\lambda^2 + \frac{\lambda}{2\beta} + M^2 - \frac{1}{4} = 0$, whose solutions are

$$\lambda = -\frac{1}{4\beta} \pm \sqrt{\frac{1}{16\beta^2} + \frac{1}{4} - M^2}. \quad (9)$$

Then, $U_0$ can be found as

$$U_0^{-4} = -\frac{\lambda}{Q\beta}\left(\frac{1}{4} + \beta^2\right). \quad (10)$$

This solution is meaningful if Eq. (9) yields real $\lambda$, and then Eq. (10) yields $U_0^4 > 0$ if $Q$ is positive, because both roots for $\lambda$, as given by Eq. (9), are negative. In turn, the condition that Eq. (9) gives real $\lambda$ implies that $\beta$ should be small enough (in other words, diffusion losses should not be too strong):

$$\beta < \frac{1}{4\sqrt{M^2 - 1/4}}, \quad (11)$$

i.e., $\beta < \frac{1}{2\sqrt{3}}$ for $M = 1$, $\beta < \frac{1}{2\sqrt{15}}$ for $M = 2$, $\beta < \frac{1}{2\sqrt{35}}$ for $M = 3$, etc.

Note also that, in the limit of $\beta = 0$ (no diffusion), the asymptotic solution (7) remains valid, with $\lambda = 0$ (i.e., the chirp vanishes), and the amplitude takes the form of $U_0^{-4} = (2Q)^{-1}(M^2 - 1/4)$.

We have numerically checked the vortex solutions with asymptotic form given by Eq. (7) under condition Eq. (11). The results show that the soliton solution indeed exists and this condition does hold for $M = 1$, but it does not hold for $M > 1$, as the single multiple vortex splits into a cluster of unitary ones, cf. Ref. [54]. Next, we seek for stable vortex solutions to Eq. (1) in a numerical form.

## 3. NUMERICAL RESULTS

The generic results produced by systematic numerical simulations of Eq. (1) may be adequately represented for values $\nu = 0.1$ and $\mu = 1$ of the parameters, which define the strength of the quintic terms in Eqs. (1) and (2). These values are fixed below. The input is a Gaussian pattern (beam, in terms of the optics model) with imprinted vorticity $M$:

$$u = u_0 \exp\left(-\frac{x^2+y^2}{2w^2}\right)\exp(iM\theta), \quad (12)$$

where $u_0$ and $w$ represent the amplitude and width of the input, respectively. The split-step fast-Fourier-transform method was adopted to simulate the evolution of input in the framework of Eq. (1), with fixed $u_0 = 1.2$ and $w = 1.2$. Note that the input taken as per Eq. (12) generates vortex solitons in the direct simulations, even if it does not include the factor given by Eq. (7), which is a property of stationary solitons. To avoid the singularity caused by the AC term in Eq. (1), assuming that the field does not exactly vanish, in the following numerical simulations we add a perturbation to the input field. The perturbation is much smaller than the main part of the input, and does not produce any conspicuous effect on numerical results.

First, we have collected the results produced by varying $M$ and the AC coefficient $Q$. Different propagation scenarios in the respective parameter domain are presented in Fig. 1(a), fixing $\beta = 0.2$ and parameters $\delta = 0.3$ and $\varepsilon = 2$ in Eqs. (2) and (3). In region B in Fig. 1(a), the input Gaussian swells into a spreading pattern when AC coefficient $Q$ is too small to stabilize localized states, as seen in Fig. 1(b). Next, if $Q$ takes values in region C in Fig. 1(a), the system is trying to generate multi-vortex patterns, which also expand in the course of the propagation (Fig. 1(b)). In region $D$ in Fig. 1(a), the most essential result is observed, *viz.*, the formation of a stable vortex cluster (Fig. 1(d)), with the number of constituent vortices equal to vorticity $M$. The cluster begins to expand slowly with the increase of the propagation distance, but then it keeps an almost constant shape when the propagation distance exceeds a certain value. The initial profile of the input evolves into a vortex triangle under the action of the embedded angular momentum and the growing modulation amplitude driven by the nonlinearity. Effectively, the vortices are produced by the centrifugal force induced by the angular momentum. In the course of the propagation, the multi-vortex cluster keeps initial counterclockwise orientation, with the pivot of the vortex triangle fixed at the center, while the triangle rotates in the opposite clockwise direction. When $M=0$, no vortices are produced, irrespective of the value of the AC coefficient, $Q$. For still larger values of $Q$, the input decays due to strong losses (Fig. 1(e)). Note that the boundaries between different regions in Fig. 1(a) weakly depend on the input's vorticity, $M$. On the other hand, numerical simulations of Eq. (1) with $Q < 0$ leads to quick destruction of the modes, because of the strong repulsion induced by the AC term in this form. For comparison, similar propagation scenarios for $M=1$ are shown in Fig. 2.

Next we address the influence of the linear-loss and AC coefficients $\delta$ and $Q$ on the propagation of the vortical field by fixing $M=5$. The same set of four distinct propagation regimes as in Fig. 1 are identified, for this case, in Fig. 3(a). Note that region D, populated by stable vortex solitons, is narrower than the others, quickly shrinking with the increase of $\delta$ and virtually disappearing, due to the strong loss, at $\delta > 0.5$.

The influence of the effective diffusion coefficient $\beta$ (see Eq. (3)) and AC parameter $Q$ on the propagation is shown in Fig. 4. An essential observation is that the range D of the stable vortex solitons expands, while the instability domain C shrinks, with the increase of $\beta$.

Finally, in Fig. 5 we display domains of the same types of the evolution of the Gaussian input, as above, in the parameter plane of the cubic-gain and AC coefficients, ($Q,\varepsilon$). It is observed that, with the increase of $\varepsilon$, the regions of instability and stability of the vortex solitons gradually rise to a maximum position with respect to $Q$, and then gradually descend. For the smaller diffusion coefficient, $\beta = 0.25$, the instability region is broader than the stable one (see the regions C and D in Fig. 5(a)), while for larger diffusion coefficient, $\beta = 0.3$, the stability area is broader than the unstable one (see the regions C and D in Fig. 5(b)).

## 4. CONCLUSION

We have investigated the evolution of input Gaussian beams with imprinted angular momentum in the framework of the two-dimensional cubic-quintic complex Ginzburg-Landau (CGL) model including the effective uniform diffusion and the anti-cubic (AC) nonlinearity. A novel asymptotic form for the dissipative solitons at $r \to 0$, dominated by the AC term, is found analytically. Systematic numerical simulations reveal the generation of stable dissipative solitons in the form of multi-vortex clusters. Domains in the parameter space of the model, which support different outcomes of the beam evolution, have been identified, varying the coefficient of the AC nonlinearity together with the parameters of the dissipative part of the CGL equation.

**Acknowledgments:** This work was supported by the National Natural Science Foundations of China (Grants Nos. 61675001 and 11774068), the Guangdong Province Nature Foundation of China (Grant No. 2017A030311025), and the Guangdong Province Education Department Foundation of China (Grant No. 2014KZDXM059).

# Figures

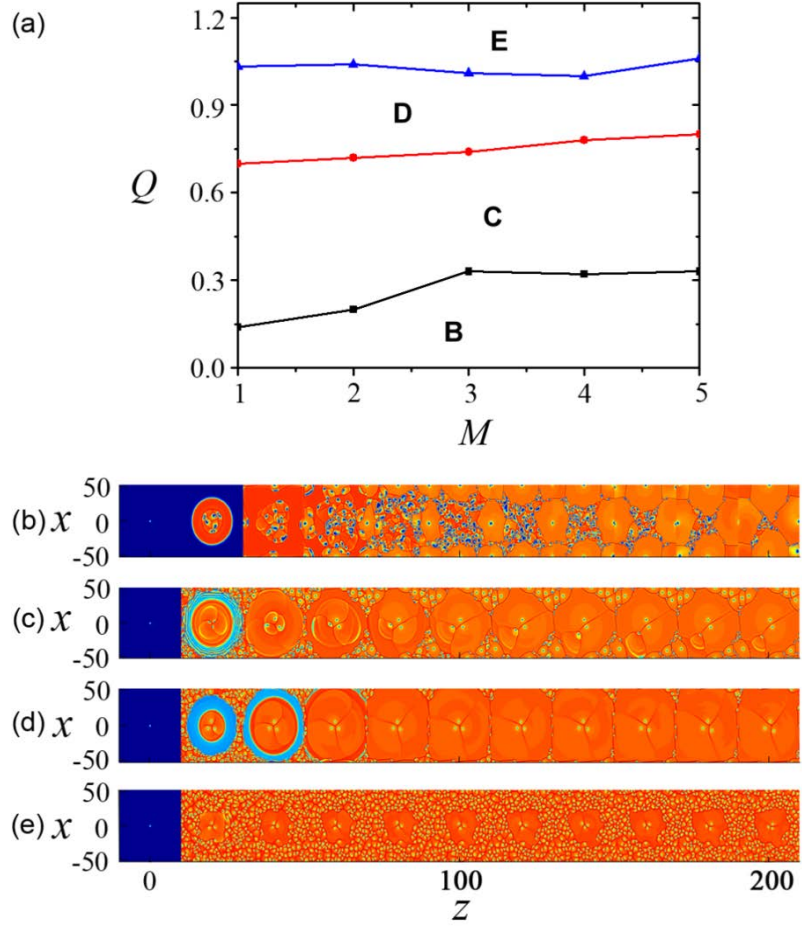

FIG. 1. (Color online) (a) Domains of different evolution scenarios for Gaussian input (12) carrying vorticity $M$ in the plane of $(Q, M)$, while other parameters in Eqs. (1)-(3) are fixed as $\beta = 0.2, \delta = 0.3$, and $\varepsilon = 2$. Region B: spatial spreading of the input beam due to energy excessive gain; C: formation of unstable vortex solitons; D: the generation of stable vortex clusters; E: decay of the input in the overdamped setting. (b) An example of the blowup for $M=3$ and $Q=0.1$, corresponding to region B in panel (a). (c) The generation of unstable vortex solitons for $M=3$ and $Q=0.4$, corresponding to region C. (d) The stable propagation of the vortex-cluster soliton with $M=3$ and $Q=0.8$, corresponding to region D. (e) The input decays for $M=3$ and $Q=1.2$.

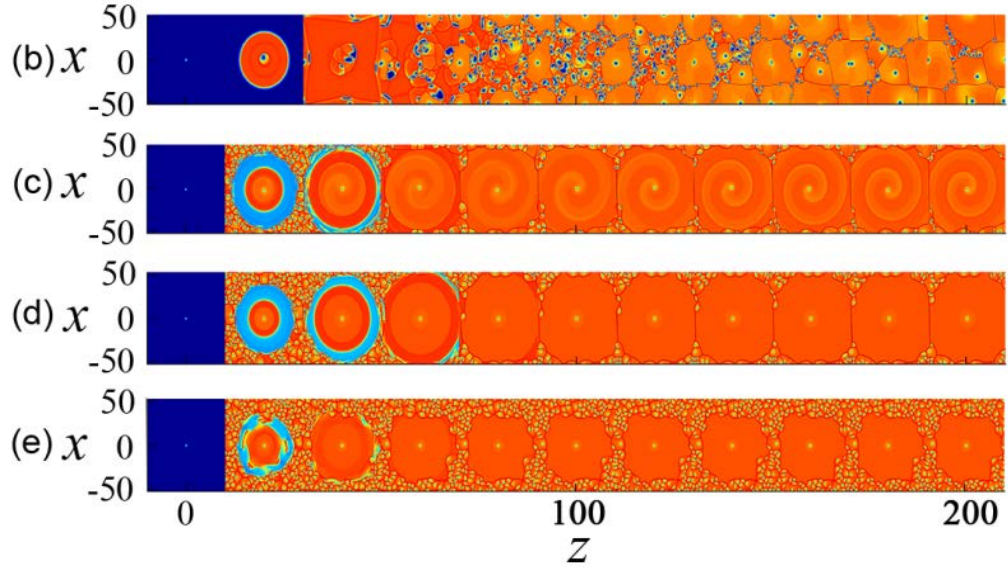

FIG. 2 (Color online) The same examples as in Fig. 1 but for $M=1$. (b) An example of the blowup for $Q=0.1$, corresponding to region B in panel (a) of Fig. 1. (c) The generation of unstable vortex solitons for $Q=0.6$, corresponding to region C in panel (a) of Fig. 1. (d) The stable propagation of the vortex-cluster soliton with $Q=0.9$, corresponding to region D in panel (a) of Fig. 1. (e) The input decays for $Q=1.1$.

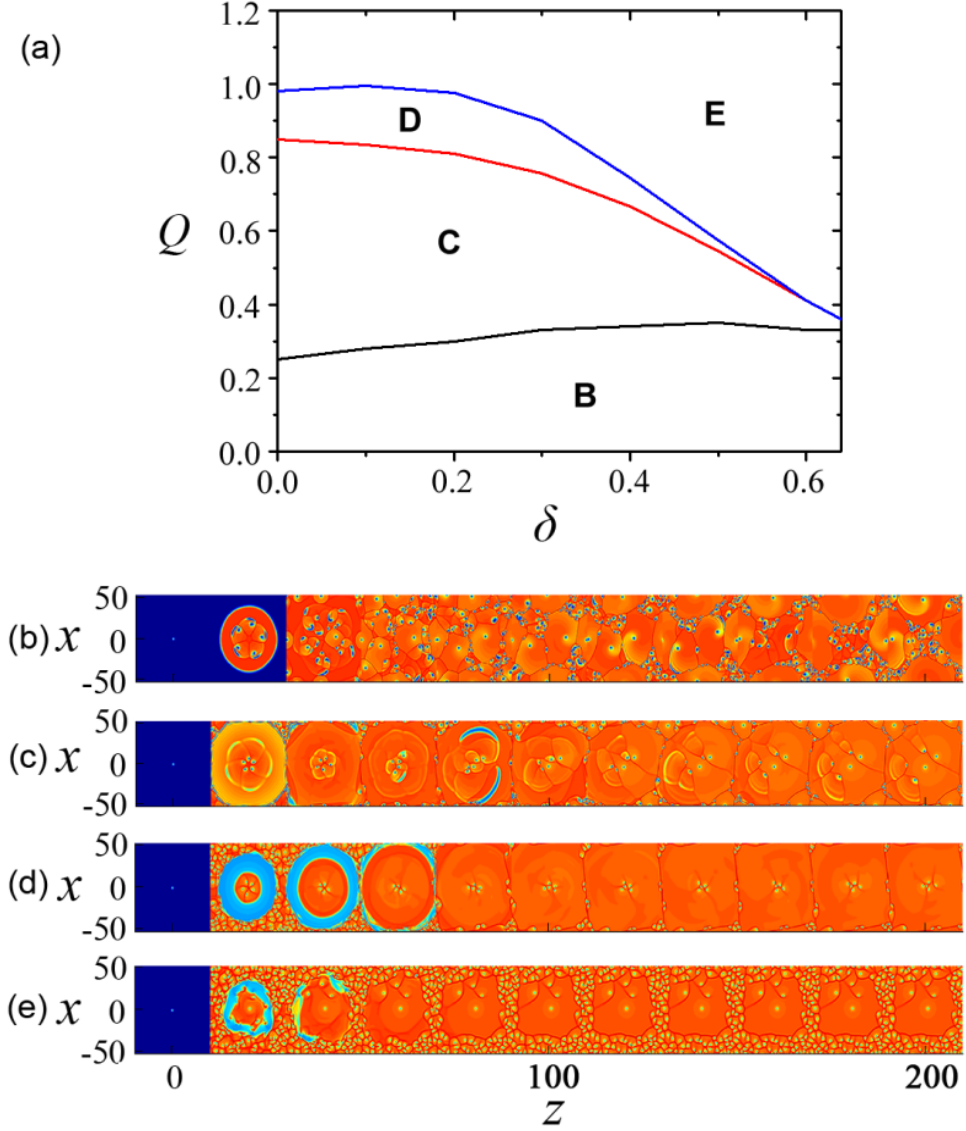

FIG. 3. (Color online) Domains of different evolution scenarios for the input Gaussian vortex in the plane of the AC and linear-loss coefficients, $Q$ and $\delta$ for fixed vorticity, $M$=5. The other fixed parameters are $\beta = 0.2$ and $\varepsilon = 2$. Regions B, C, D, and E have the same meaning as in Fig. 1. (b) An example of the beam spreading for $\delta$=0.1 and $Q$=0.1, corresponding to region B in panel (a). (c) The generation of an unstable vortex soliton for $\delta$=0.2 and $Q$=0.4, corresponding to region C. (d) The stable propagation of the vortex-cluster soliton for $\delta$=0.3 and $Q$=0.9, corresponding to region D. (e) The input decays at $\delta$=0.5 and $Q$=1.1.

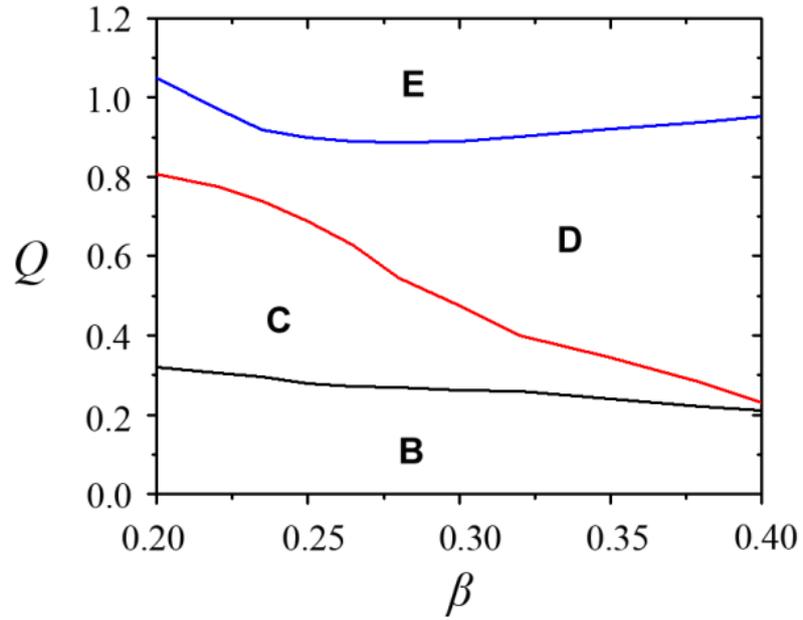

FIG.4. (Color online) The same domains of different evolution scenarios as in Figs. 1(a) and 3(a) for the input Gaussian vortex beam in the plane of the diffusion and AC coefficients, $\beta$ and $Q$ for fixed $M = 5$. The other parameters are $\delta=0.3$ and $\varepsilon = 2$.

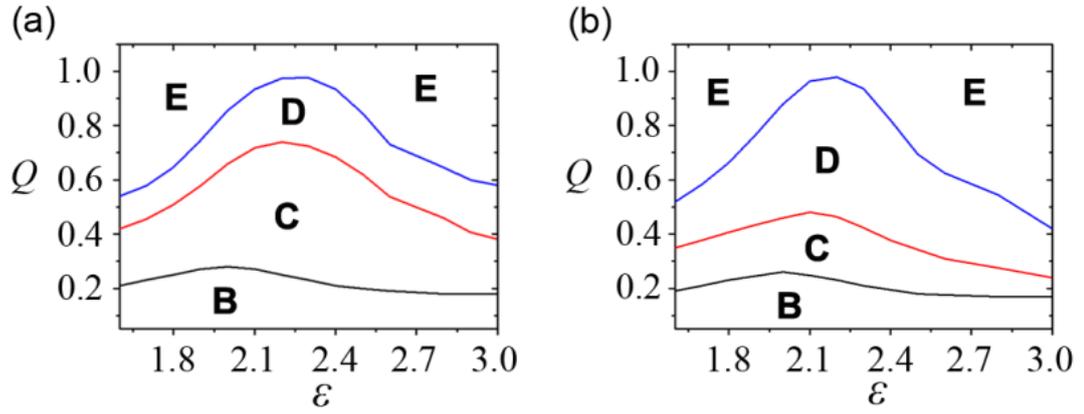

FIG. 5. (Color online) Domains of different evolution scenarios for the input Gaussian vortex beam in the plane of the cubic gain and AC coefficient, $(\varepsilon, Q)$, for fixed $M=5$ and $\delta=0.3$. The diffusion coefficient is $\beta=0.25$ in (a), and $\beta=0.3$ in (b). The meaning of domains B, C, D, and E is the same as in the above figures.

# Figures captions

FIG. 1. (Color online) (a) Domains of different evolution scenarios for Gaussian input (12) carrying vorticity $M$ in the plane of ($Q$, $M$), while other parameters in Eqs. (1)-(3) are fixed as $\beta = 0.2, \delta = 0.3$, and $\varepsilon = 2$. Region B: spatial spreading of the input beam due to energy excessive gain; C: formation of unstable vortex solitons; D: the generation of stable vortex clusters; E: decay of the input in the overdamped setting. (b) An example of the blowup for $M$=3 and $Q$=0.1, corresponding to region B in panel (a). (c) The generation of unstable vortex solitons for $M$=3 and $Q$=0.4, corresponding to region C. (d) The stable propagation of the vortex-cluster soliton with $M$=3 and $Q$=0.8, corresponding to region D. (e) The input decays for $M$=3 and $Q$=1.2.

FIG. 2 (Color online) The same examples as in Fig.1 but for $M$=1. (b) An example of the blowup for $Q$=0.1, corresponding to region B in panel (a) of Fig. 1. (c) The generation of unstable vortex solitons for $Q$=0.6, corresponding to region C in panel (a) of Fig. 1. (d) The stable propagation of the vortex-cluster soliton with $Q$=0.9, corresponding to region D in panel (a) of Fig. 1. (e) The input decays for $Q$=1.1.

FIG. 3. (Color online) Domains of different evolution scenarios for the input Gaussian vortex in the plane of the AC and linear-loss coefficients, $Q$ and $\delta$ for fixed vorticity, $M$=5. The other fixed parameters are $\beta = 0.2$ and $\varepsilon = 2$. Regions B, C, D, and E have the same meaning as in Fig. 1. (b) An example of the beam spreading for $\delta$=0.1 and $Q$=0.1, corresponding to region B in panel (a). (c) The generation of an unstable vortex soliton for $\delta$=0.2 and $Q$=0.4, corresponding to region C. (d) The stable propagation of the vortex-cluster soliton for $\delta$=0.3 and $Q$=0.9, corresponding to region D. (e) The input decays at $\delta$=0.5 and $Q$=1.1.

FIG. 4. (Color online) The same domains of different evolution scenarios as in Figs. 1(a) and 3(a) for the input Gaussian vortex beam in the plane of the diffusion and AC coefficients, $\beta$ and $Q$ for fixed $M = 5$. The other parameters are $\delta$=0.3 and $\varepsilon = 2$.

FIG. 5. (Color online) Domains of different evolution scenarios for the input Gaussian vortex beam in the plane of the cubic gain and AC coefficient, $(\varepsilon, Q)$, for fixed $M=5$ and $\delta=0.3$. The diffusion coefficient is $\beta=0.25$ in (a), and $\beta=0.3$ in (b). The meaning of domains B, C, D, and E is the same as in the above figures.